# Specific and non-specific hybridization of oligonucleotide probes on microarrays

Hans Binder and Stephan Preibisch

Interdisciplinary Centre for Bioinformatics, University of Leipzig

Interdisciplinary Centre for Bioinformatics of Leipzig University, D-4107 Leipzig, Haertelstr. 16-18, binder@izbi.uni-leipzig.de, fax: ++49-341-9716679

**Abstract**

Gene expression analysis by means of microarrays is based on the sequence specific binding of mRNA to DNA oligonucleotide probes and its measurement using fluorescent labels. The binding of RNA fragments involving other sequences than the intended target is problematic because it adds a "chemical background" to the signal, which is not related to the expression degree of the target gene. The paper presents a molecular signature of specific and non-specific hybridization with potential consequences for gene expression analysis. We analyzed the signal intensities of perfect match (PM) and mismatch (MM) probes of GeneChip microarrays to specify the effect of specific and non-specific hybridization. We found that these events give rise to different relations between the PM and MM intensities as function of the middle base of the PM, namely a triplet-like (C>G≈T>A>0) and a duplet-like (C≈T>0>G≈A) pattern of the PM-MM log-intensity difference upon binding of specific and non-specific RNA fragments, respectively. The systematic behaviour of the intensity difference can be rationalized on the level of base pairings of DNA/RNA oligonucleotide duplexes in the middle of the probe sequence. Non-specific binding is characterized by the reversal of the central Watson Crick (WC) pairing for each PM/MM probe pair, whereas specific binding refers to the combination of a WC and a self complementary (SC) pairing in PM and MM probes, respectively. The Gibbs free energy contribution of WC pairs to duplex stability is asymmetric for purines and pyrimidines and decreases according to $C > G \approx T > A$. SC pairings on the average only weakly contribute to duplex stability. The intensity of complementary MM introduces a systematic source of variation which decreases the precision of expression measures based on the MM intensities.

**Introduction**

Understanding of factors affecting the transcription of genetic information into the proteome level is one of the major challenges in the context of systems biology and molecular medicine. It requires new high-throughput techniques to analyse the activity of a large number of potentially important genes. The high-density-oligo-nucleotide-array (HDONA) technology enables to estimate the expression degree of thousands of genes in particular cells or tissues at once by the measurement of the abundance of the respective messenger RNA. This method is based on both, the sequence specific binding (hybridization) of the "target" RNA to complementary DNA oligonucleotide probes and the fluorescence labelling and detection of probe-bound RNA transcripts as well. For example, up to one million probes of different sequences referring to 20.000 – 45.0000 different genes are attached to typical microarrays of the GeneChip type in spots of a few $\mu m^2$ per probe (Lipshutz et al., 1999).

The integral fluorescence intensity per probe array is directly related to the amount of bound RNA, which in turn serves as a measure of the target RNA concentration in the studied sample solution. It represents a mixture of RNA fragments with a wide distribution of different sequences. A considerable amount of RNA fragments consequently involves other sequences than the intended target of a selected probe. Unfortunately also these non-specific transcripts can possess a non negligible affinity for duplex formation with the probes. In other words, duplex formation between RNA transcripts and the DNA probes partially lacks specificity in terms of complementary Watson Crick (WC) base pairings. This non-specific hybridization is problematic for chip analysis because it adds a "chemical" background intensity, which is not related to the expression degree of the target gene.

One experimental option to deal with this problem is the pairwise design of each probe sequence on Affymetrix© GeneChip micoarrays (Affymetrix, 2001a). The sequence of the 25-meric so-called perfect match (PM) probe is taken from the target gene and thus it is complementary to a sequence length of 25 nucleotide bases in the transcribed target RNA. On the other hand, the so-called mismatch probe (MM) is identical with the PM probe except the base in the middle of the sequence, which is replaced by its complement to prevent specific hybridization, i.e. the binding of the target RNA. This way, the MM probe intends to measure the amount of non-specific hybridization, and thus to provide a correction of the PM intensity for the chemical background. In addition, a certain number (usually 11-20) of PM/MM probe pairs taken from different regions of the same gene form a so-called probe set to get several estimates of its expression degree and thus to improve the reliability of the method.

The idea behind the correction using mismatches is based on the assumption that non-specific binding is identical for PM and MM probes, i.e., non-specific transcripts do not see the letter change in the middle of the sequence. It is further assumed in accordance with conventional hybridization theory that the mismatch strongly reduces the affinity of target binding to the MM, and thus specific transcripts see the change of the middle letter (Li and Wong, 2001a; Li and Wong, 2001b). These assumptions predict a systematically equal or higher intensity of the PM compared with that of the MM, $I^{PM} \geq I^{MM}$, given that the fluorescence response per bound transcript is identical for PM and MM and for specific and non-specific hybridization as well.

Chip analyses however show that a fair number of MM probes posses a larger fluorescence intensity than their PM counterpart (Naef et al., 2002). It was concluded that conventional hybridization theory is simply inadequate, and particularly, that the basic mechanism of MM hybridization is not understood yet. As a consequence many algorithms of gene expression analysis simply ignore MM intensity data (see, e.g., (Zhou and Abagyan, 2002) and (Irizarry et al., 2003) for an overview) or the MM are considered in an empirical fashion to exclude "bad" probes from the analysis (Affymetrix, 2001a; Affymetrix, 2001b). Other publications discuss nonlinearities in the probe responses and sequence effects in the behavior of matched and mismatched probes showing that the hybridization on microarrays is apparently a complex phenomenon, which is governed by an intricate interplay between several effects such as the stability of RNA/DNA duplexes, binding and saturation, surface electrostatics and diffusion, fluorescence emission and non equilibrium thermodynamics (Bhanot et al., 2003; Chan et al., 1995; Chudin et al., 2001; Dimitrov and Zuker, 2004; Halperin et al., 2004; Hekstra et al., 2003; Held et al., 2003; Naef et al., 2002; Naef and Magnasco, 2003; Vainrub and Pettitt, 2002; Zhang et al., 2003).

The "riddle of bright MM" was apparently solved by Naef et al. (Naef and Magnasco, 2003) who showed that the difference between the PM and MM intensities strongly correlates with the middle base at position k=13 of the 25-meric probe. For probe pairs with double ringed pyrimidines (C, T) in the middle of the PM sequence one finds a preference for "bright" PM, $I^{PM} > I^{MM}$. In contrast, for purines (G, A) the relation reverses with the tendency for "bright" MM. The interpretation in terms of probe-target duplexes suggests that single-ringed pyrimidines form stronger self complementary (SC) base pairings (i.e., C•c* and T•u*, lower case letters refer to the RNA, the asterisk denotes fluorescent labeling and mismatched base pairings are underlined) compared with the respective WC pairs (C•g and T•a) owing to steric effects and labelling (Naef and Magnasco, 2003).

On the other hand it is well accepted that SC pairs between oligonucleotides in solution are considerably weaker than WC pairs (Peyret et al., 1999; Sugimoto et al., 2000). Studies on the hybridization of mismatched probes on different microarray types reveal agreement with solution data (Dorris et al., 2003; Ramakrishnan et al., 2002). Hence, the postulated SC base pair interactions on GeneChip microarrays contradict "conventional" hybridization properties of oligonucleotides in solution and also on microarrays. The fundamentally different behavior of GeneChip probes (the so-called "riddle of bright MM") is intriguing but also strange because it seems to violate conventional hybridization rules.

The accurate interpretation of microarray intensity data in terms of the expression degree remains a significant challenge, which requires the understanding of the hybridization behavior on the level of base-pair interactions. The present publication aims at examining the validity of the basic rules of DNA/RNA hybridization in solution for hybridization on HDONA microarrays and at extracting a molecular signature to discriminate specific and non-specific hybridization on the level of base pairings in DNA/RNA duplexes.

## Chip data

The classification of the probes according to perfect-matched and mismatched pairings of the middle base refers to specific duplexes of the PM and MM probes with the complementary sequence of the respective target RNA. Consequently the interpretation of MM intensity data in terms of SC base pairings assumes exclusively specific hybridization of the MM probes, a condition which is usually not realized. The present study therefore separates specific and non-specific hybridization using a special calibration data set to analyze the PM and MM probe intensities in terms of base pair interactions in RNA/DNA duplexes on microarrays.

Particularly, the microarray intensity data of PM and MM probes, $I_p^{PM}$ and $I_p^{MM}$ (p is the probe no.), respectively, are taken from the Affymetrix' human genome HG U133 Latin Square (HG U133-LS) data set available at http://www.affymetrix.com/support/technical/sample_data/datasets.affx. The HG U133-LS experiment considers transcripts of 42 genes (42x11=462 different probes). They are titrated ("spiked") onto 14 different arrays at 14 concentrations corresponding to all cyclic permutations in a complex human background extracted from a HeLa cell line not containing the spikes. This way one gets the relation between the probe intensities and the respective ("spiked-in") concentration of specific RNA. Each condition was realized in triplicate. PM and MM intensities are background corrected using the algorithm provided by MAS 5.0 (Affymetrix, 2001a; Affymetrix, 2001b).

## Results

**The effect of "bright" MM probe intensities is related to non-specific hybridization**

More than 30% of all probe pairs of Affymetrix$^©$ GeneChips are characterized by "bright" mismatched MM probes, which show a higher intensity and thus a stronger affinity for duplex formation with RNA fragments than the respective perfect matched PM probes although the middle base in the MM does not match the target sequence in terms of Watson-Crick (WC) pairs (Naef et al., 2002). To analyse this effect as a function of the relative amount of specific transcripts we plot the log-intensity difference, $\log I_p^{PM-MM} \equiv \log I_p^{PM} - \log I_p^{MM}$, of all spiked-in probes pairs at all available concentrations of specific transcripts (0 pM $\leq c_{RNA}^S \leq$ 512 pM) as function of the set-averaged mean log intensity, $<\log I_p^{PM+MM}>_{set} \equiv 0.5<(\log I_p^{PM} + \log I_p^{MM})>_{set}$, which serves as an empirical measure of the concentration of specific transcript (Binder et al., 2004b) (see Fig. 1). Note that the usually eleven probes of each set refer to one target gene and thus to specific RNA fragments of one concentration.

We used this simple parameter instead of other estimates of the relative transcript concentration (see ref. (Irizarry et al., 2003) for an overview) because (i) it can be calculated for single chips, i.e., it is not based on the comparison of the probe intensities of several chips; (ii) the computation of $<\log I_p^{PM+MM}>_{set}$ is rather simple; and (iii) it includes no correction for the chemical background, the identification of which is one goal of the present work. The $\log I_p^{PM-MM}$ data are separately re-plotted for three selected spiked-in concentrations in the lower panel of Fig. 1. It shows that the concentration of specific transcripts well correlates with the set averaged log-intensity, $<\log I_p^{PM+MM}>_{set}$, which however spreads with an uncertainty of $\delta<\log I^{PM+MM}> \approx \pm 0.5$ for each spiked-in concentration.

The lower part of Fig. 1 clearly reveals that the PM-MM log-intensity difference increases with increasing amounts of specific transcripts. In particular, the cloud of the $\log I_p^{PM-MM}$ data markedly shifts upwards with increasing $c_{RNA}^S$. The parallel increase of the mean intensity difference averaged over all spiked-in probes of one concentration, $<\log I_p^{PM-MM}>_{c=const}$, clearly reflects this trend (see Fig. 2). The onset of saturation gives rise to a maximum of the averaged log intensity difference at higher concentrations and the decrease of $<\log I_p^{PM-MM}>_{c=const}$ with further increasing $c_{RNA}^S$. The set-averaged intensity difference, $<\log I_p^{PM-MM}>_{set}$ (open symbols in Fig. 1), and especially the mean log intensity difference of all probes of one spiked-in concentration, $<\log I_p^{PM-MM}>_{c=const}$ (Fig. 2, panel above), more clearly indicate this trend.

For a more detailed analysis we also calculated the fraction of probe pairs with "bright" MM, $f(MM>PM)_{c=const}=N(MM>PM)_{c=const}/N_{total}^{sp-in}$ (see Fig. 2, panel below, $N_{total}^{sp-in}=462$ is the total number of spiked-in probes and $N(MM>PM)_{c=const}$ is the number of probes meeting the condition of bright MM, $\log I_p^{PM-MM} < 0$) for each spiked-in concentration to characterize the intensity relation between the PM and MM as a function of $c_{RNA}^S$, the concentration of specific spiked-in transcripts. The fraction of probe pairs with "bright" MM decreases from $f(MM>PM) \approx 0.43$ in the absence of specific transcripts to values smaller than 0.05 at $c_{RNA}^S > 100$ pM. Hence the intensity of almost all 462 PM probes referring to the spiked-in transcripts exceeds the intensity of the respective MM if the RNA binding is dominated by specific hybridization.

**Figs. 1 and 2**

In the absence of specific hybridization nearly one half of all spiked-in probe pairs gives rise to "bright" MM. Owing to this effect more than 20% of the spiked-in probe sets are characterized by a larger set averaged MM intensity compared with the respective PM value (i.e., $<\log I^{PM-MM}>_{set} < 0$, see also the open circles in Fig. 1, which show the set averaged log-intensity differences of the spiked-in probes). The respective fraction of probe sets of "bright MM", $f^{set}(MM>PM)$, more steeply decreases with increasing concentration of specific transcripts than the overall fraction of single "bright" MM probes, $f(MM>PM)$ (see triangles in the lower panel of Fig. 2). This difference can be simply explained by means of the binominal distribution $B(n,N,p) = \binom{N}{n} p^n (1-p)^{N-n}$, where $p = f(MM>PM)$ is the probability to find a probe pair with "bright MM". It predicts the probability that n = N(MM>PM) probe pairs meet the condition $I^{MM}>I^{PM}$ within an independent set of N = $N_{set}$ probe pairs, if one assumes that the sequence specific affinities of the probes are randomly distributed among the probe sets (see below) and that the PM and MM log-intensities are equally distributed about the set averages. Then, the fraction of "bright MM probe sets" is to a good approximation given by the probability that more than 50% of the probe pairs of the set possess bright MM, i.e. $f^{set}(MM>PM) \approx \sum_{n=n(\min)}^{N} B(n,N,p)$ with n(min)≈0.5·$N_{set}$. Figure 2 shows that the experimental data are well compatible with n(min) = 6 – 7 (compare the triangles with the curves "6" and "7") in agreement with the prediction.

**Fig. 3**

To generalize these results we calculate the fraction of "bright" MM and the mean log-intensity difference for all 250.000 probes of a HG U133 chip (see Fig. 3). The respective running averages of f(MM<PM) and of <log$I_p^{PM-MM}$> show virtually the same features as the respective curves of the spiked-in genes (compare with Fig. 2). Note that the x-axes in both figures, the concentration in Fig. 2 and mean intensity in Fig. 3, scale non-linearly each to another. For example the plateau of f(MM<PM) and of <log$I_p^{PM-MM}$> at small intensity values <log$I_p^{PM+MM}$>$_{set}$ < 1.8 (see Fig. 3) can be mainly attributed to non-specifically hybridized probes referring to the smallest concentration values, $c_{RNA}^S$ < 0.2 pM, in Fig. 2. We conclude that the scaling of the probes with the set-averaged mean log-intensities indeed reflects essential properties of the concentration dependence as suggested previously (Binder et al., 2004b).

**Figs. 4 and 5**

**The effect of "bright" MM probe intensities is related to the middle base**

It was previously found that the effect of "bright" MM and thus the difference between the PM and MM intensities strongly correlates with the middle base at position k=13 of the 25-meric probe if one considers all probes of the chip (Naef and Magnasco, 2003). For probe pairs with double ringed pyrimidines (C, T) in the middle of the PM sequence one finds a preference for "bright" PM, $I^{PM} > I^{MM}$. In contrast, for purines (G, A) the relation reverses with the tendency for "bright" MM. The analysis in terms of probe sensitivities (see below) reveals a similar result (Binder et al., 2004a; Binder et al., 2004b).

To shed light into the effect of specific and non-specific hybridization on the observed bias due to the middle base we separately plot the intensity difference, log$I_p^{PM-MM}|_B$, for all probe pairs of the chip possessing a common middle base B = A, T, G and C of the PM probe (see the upper panel of Figs. 4 and 5). The respective data cloud systematically shifts upwards for pyrimidines, B=C and T, and downwards for purines, B=G and A, as expected. The respective fraction of "bright" MM, $f_B(MM<PM)$, and the mean log-intensity difference,

$$\log I_B^{PM-MM} \equiv \log I_B^{PM} - \log I_{Bc}^{MM} = \left\langle \log I_p^{PM}|_B \right\rangle - \left\langle \log I_p^{MM}|_{Bc} \right\rangle \tag{1}$$

of probes with middle bases B=A,T,G,C (PM) and its complementary base $B^c$=T,A,C,G (MM) considerably deviate from the overall mean over all probes (compare lower panel of Figs. 4 and 5 with Fig. 3). In probe pairs with B=A, G more than 60% of the MM are "bright" in the "plateau region" of $f_B$(MM<PM), which refers to hybridization with a dominating fraction of non-specific transcripts. In contrast, only about 20% of the probe pairs possess "bright" MM for B=T, C in the respective range of small mean intensities.

### Figs. 6 and 7

In Figs. 6 and 7 we plot the middle base-specific fraction of bright MM, $f_B$(MM<PM) (panel below), and the respective mean PM-MM difference, $\log I_B^{PM-MM}$ (panel above), for comparison of the chip averages (Fig. 6) with the respective averages over the spiked-in probes (Fig. 7). Both kinds of data show essentially identical properties indicating, (i) that the whole ensemble of probes behaves similarly compared with the reduced ensemble of spiked-in probes, and (ii) that the concentration dependence of the specific transcripts transforms into the scale of the set-averaged intensity to a good approximation (see above).

The mean difference of log intensities, $\log I_B^{PM-MM}$, is negative for the middle bases B=A and G and clearly positive for T and C with values, which obey a duplet-like pattern according to the relation C≈T>0>G≈A in the limit of non-specific hybridization. The $\log I_B^{PM-MM}$-curves split into four different courses in the intermediate intensity range according to C>T>G>A>0, and finally the G and T curves merge together giving rise to a triplet-like pattern with C>T≈G>A>0 at high mean intensities, i.e., in the limit of dominating specific hybridization. Hence, the systematic shift between the PM-MM intensity differences is clearly affected by the relative amount of specific hybridization indicating that specific and non-specific transcripts bind differently to probes with a certain middle base.

The slightly smaller fraction of bright MM for B=A,G in the full data set compared with the spiked-in set at small abscissa values can be attributed to the fact that a small amount of specific transcripts also contributes to the respective averages in the limit of small abscissa-values of the mean intensity.

**Middle-base averaged probe sensitivity**

In a next step we transform the log-intensity difference referring to one middle base into a relative scale with respect to the total mean over all spiked-in probes of one concentration ($<...>_{c=const}$) by means of

$$Y_B^P = \log I_B^P - \left\langle \log I_p^P \right\rangle_{c=const} \quad , \quad P = PM - MM \qquad (2)$$

Equation 2 defines the middle base related sensitivity difference between perfect matched and mismatched oligonucleotide probes. Note that the sensitivity characterizes the ability of a probe to detect a certain amount of RNA (Binder et al., 2004b). It depends on the binding affinity (i.e. the binding "strength" for duplex formation with the target) and on the fluorescence yield (which is related to the intensity per bound transcript, i.e., to the number of fluorescence labels attached to the RNA sequence) of the relevant RNA transcripts. The middle-base related sensitivity given by Eq. 2 is expected to filter out the systematic effect of the respective middle base on the PM-MM log-intensity difference. Figure 8 shows the respective sensitivity data which are derived from the Latin square experiment as a function of the specific transcript concentration of the spiked-in probes, $c_{RNA}^{S}$ (see also Fig. 7).

**Figs. 8 and 9**

In the limit of dominating non-specific hybridization at small $c_{RNA}^{S}$ values one obtains a duplet-like relation between the data, $Y_C^{PM-MM,NS} \approx Y_T^{PM-MM,NS} \approx -Y_G^{PM-MM,NS} \approx -Y_A^{PM-MM,NS}$. With increasing $c_{RNA}^{S}$ the absolute sensitivity values for B = G,T progressively decrease and virtually merge in the limit of dominating specific hybridization revealing a triplet-like pattern according to $Y_C^{PM-MM,S} \approx -Y_A^{PM-MM,S} > Y_T^{PM-MM,S} \approx Y_G^{PM-MM,S}$. The slight decrease of the absolute values of $Y_C^{PM-MM,S}$ and of $Y_A^{PM-MM,S}$ with increasing specific transcript concentrations $c_{RNA}^{S}$ presumably reflects saturation (see Fig.8 and ref. (Binder et al., 2004b)).

**Positional dependent single base (SB) model**

To further specify the effect of each single base along the probe sequences on the observed sensitivity difference we used a simple model, which approximates the sensitivity of P = PM, MM probes,

$$Y_p^{P,h} = \log I_p^P - \langle \log I_p^P \rangle_{p \in \Sigma} \quad with \quad h = NS, S \quad and \quad \Sigma = \Sigma^h \quad , \tag{3}$$

by a sum of base and positional dependent sensitivity terms,

$$Y_p^{P,SB} = \sum_{k=1}^{25} \sum_{B=A,T,G,C} \sigma_k^P(B) \cdot \left( \delta(B, \xi_{p,k}^P) - f_k^\Sigma(B) \right) \quad , \quad P = PM, MM \quad . \tag{4}$$

The considered probes (index p) were taken from a subset of all probes on the chip, $\Sigma^h$, which refers predominantly to non-specifically (h=NS) and specifically (h=S) hybridized probes (i.e., $p \in \Sigma^h$). We chose all probe sets which meet the condition $<\log I_p^{PM+MM}>_{set} < 1.8$ for the subset $\Sigma^{NS}$ and $<\log I_p^{PM+MM}>_{set} > 2.8$ for the subset $\Sigma^S$ according to the correlation between the set-averaged log-intensities and the spiked-in concentration established above. δ denotes the Kronecker delta (δ(x,y)=1 if x=y and δ(x,y)=0 if x≠y) and $f_k^\Sigma(B)$ is the fraction of base B at position k in the considered ensemble of probes, $\Sigma^h$. The nucleotide base at position k along the sequence of probe number p is denoted by $\xi_{p,k}^P$. The values of the positional dependent sensitivity terms for each base, $\sigma_k^P(B)$, were estimated by multiple linear regression of the experimental and theoretical sensitivities, $Y_p^{P,h}$ and $Y_p^{P,SB}$, respectively, using singular value decomposition for solving the obtained system of linear equations (see (Binder et al., 2005) for details).

The sensitivity profiles of the PM probes of both subsets, $\Sigma^S$ and $\Sigma^{NS}$, and of the non-specifically hybridized MM probes are very similar, i.e. $\sigma_k^{PM,S}(B) \approx \sigma_k^{PM,NS}(B) \approx \sigma_k^{MM,NS}(B)$ (see Fig. 9, panels above). In particular, the profiles for B=C, A show the typical parabola-like shape being maximum and minimum in the middle of the sequence, respectively, whereas the sensitivity terms for B=T, G change almost monotonously along the sequence with their minimum and maximum values at k=1, respectively (see also (Binder et al., 2003; Binder et al., 2005; Mei et al., 2003; Naef and Magnasco, 2003)).

The sensitivity profiles of specifically hybridized MM probes distinctly differ in the middle of the sequence from the other considered profiles for B=A, C. Namely, the absolute values of $\sigma_{13}^{MM,S}(C)$ and $\sigma_{13}^{MM,S}(A)$ markedly drop to values near zero giving rise to a "dent-like" shape of the respective curves. Note that also the sensitivity profiles of B=G, T adopt only tiny values at k=13. One can therefore assume $\sigma_{13}^{MM,S}(B) \approx 0$ for all bases B=A,T,G,C to a good approximation. In other words, there is on the average only a weak base-specific contribution from the mismatched middle base of the MM probes to the respective probe intensities in the limit of specific hybridization. On the other hand, the matched bases at the remaining sequence positions k ≠ 13 give rise to similar sensitivity profiles of the PM and MM probes in the limit of specific and non-specific hybridization as well, i.e., $\sigma_k^{PM,h}(B) \approx \sigma_k^{MM,h}(B)$ for k≠13 and h=N, NS.

For the further discussion of the positional effect on the PM-MM sensitivity difference let us rewrite the SB model for each PM/MM pair:

$$Y_p^{PM-MM,SB} = \sum_{B=A,T,G,C} \left( \sum_{k=1}^{25} \sigma_k^{PM-MM}(B) \cdot \left( \delta(B, \xi_{p,k}^P) - f_k^\Sigma(B) \right) \right) \quad \text{with}$$

$$\sigma_k^{PM-MM}(B) = \begin{cases} \sigma_k^{PM}(B) - \sigma_k^{MM}(B) & \text{for} \quad k \neq 13 \\ \sigma_{13}^{PM}(B) - \sigma_{13}^{MM}(B^c) & \end{cases}$$ 

(5)

Equation 5 takes into account that the sequences of the PM and MM probes of each pair are identical for all positions k≠13 but complementary for the middle bases at k=13. The lower panel of Fig. 9 shows the respective difference profiles. The $\sigma_k^{PM-MM}(B)$-values virtually vanish for k≠13, as expected. On the other hand, the sensitivity difference of the middle base considerably differs from zero. The $\sigma_{13}^{PM-MM}(B)$-values change in a similar fashion as the middle-base related sensitivity differences $Y_B^{PM-MM}$ with increasing amount of specific transcripts (see Fig. 8 and previous section). Namely, the difference of the sensitivity terms split into a duplet, $\sigma_{13}^{PM-MM}(C) \approx \sigma_{13}^{PM-MM}(T) \approx -\sigma_{13}^{PM-MM}(A) \approx -\sigma_{13}^{PM-MM}(G)$, in the limit of non-specific hybridization and into a triplet, $\sigma_{13}^{PM-MM}(C) \approx -\sigma_{13}^{PM-MM}(A) > \sigma_{13}^{PM-MM}(T) \approx \sigma_{13}^{PM-MM}(G)$, in the limit of specific hybridization in correspondence with the behavior of $Y_B^{PM-MM}$. The analysis of the spiked-in probes in terms of the SB model provides similar results (not shown here, see (Binder et al., 2005)).

The parallel behavior of the SB sensitivity difference of the middle base (see Eq. 5 and Fig. 9, panel below) and of the middle base averaged mean sensitivity difference (Eq. 2, see Fig. 8) is plausible because the averaging to a high degree reduces the specific effect of the bases at positions k=1-12 and 14-25. In other words, the observed variation of $Y_B^{PM-MM}$ can be mainly attributed to the middle base, i.e.

$$Y_B^{PM-MM} \approx \sigma_{13}^{PM-MM}(B)$$ 

(6)

Note that $Y_B^{PM-MM}$ and $\sigma_{13}^{PM-MM}(B)$ are the results of independent analyses where the former one simply averages out the effect of the bases at positions $k \neq 13$ in contrast to the latter method, which explicitly considers the mean effect of each base at each position.

## Discussion

**The affinity of DNA oligonucleotide probes for RNA binding**

Essentially four multiplicative factors affect the signal intensity of microarray probes: (i) the binding affinity of the particular probe for duplex formation with RNA fragments, (ii) the fluorescence yield of probe-bound RNA fragments depending on the number of labelled nucleotides in their sequence, (iii) the relative abundance of RNA fragments which potentially bind to the probe in the sample solution and (iv) a proportionality constant which considers effects due to chip fabrication (e.g. the surface density of probes), sample preparation (e.g., the total RNA concentration in the sample solution) and imaging (e.g., the sensitivity of the scanner) (Binder et al., 2004a). Effects (iii) and (iv) are common for a given gene and chip, respectively, and, thus they largely cancel out in the log-intensity difference, $\log I_p^{PM-MM}$, of each PM/MM probe pair. The sequences of the PM and MM probes differ only with respect to their middle base. Consequently, sequence specific effects (i) and (ii) are reduced in the log-intensity difference, $\log I_p^{PM-MM}$, compared with the individual intensity values, $\log I_p^{PM}$ and $\log I_p^{MM}$. In particular, the amount of labelling is either equal or it differs by only one labelled base if one compares the specific and non-specific duplexes of the PM with that of the MM probes, respectively. We therefore neglect the effect of labelling in the following considerations. Finally, the averaging over all probe pairs with a certain middle base according to Eq. 1 largely decreases sequence-specific effects due to base positions $k=1...12$ and $14...25$ of the 25-meric probes (Binder et al., 2004a).

Hence the middle base related log-intensity difference of a PM/MM probe pair (Eq. 1) is expected to reflect the mean effect of changing base B by its complementary base $B^c$ in the middle of oligonucleotide probes upon hybridization on GeneChip microarrays. Note that the log-intensity difference is given to a good approximation by (see (Binder et al., 2004a; Binder et al., 2004b))

$$\log I_B^{PM-MM} \approx \log K_B^{PM-MM} - \log S_B^{PM-MM}$$

with

$$\log K_B^{PM-MM} = \log \left\{ \frac{c_{RNA}^S \cdot K_B^{PM,S} + c_{RNA}^{NS} \cdot K_B^{PM,NS}}{c_{RNA}^S \cdot K_{B^c}^{MM,S} + c_{RNA}^{NS} \cdot K_{B^c}^{MM,NS}} \right\} \quad and \quad (7)$$

$$\log S_B^{PM-MM} = \log \left\{ \frac{1 + \left( c_{RNA}^S \cdot K_B^{PM,S} + c_{RNA}^{NS} \cdot K_B^{PM,NS} \right)}{1 + \left( c_{RNA}^S \cdot K_B^{MM,S} + c_{RNA}^{NS} \cdot K_B^{MM,NS} \right)} \right\}$$

where $K_B^{P,h}$ denotes the effective binding constant of the P=PM (and MM) probe with middle letter B (and $B^c$) for association with specific (h=S) and non-specific transcripts (h=NS), respectively (see also text which follows). Note that the $K_B^{P,h}$ are effective, i.e. mean values averaged over the respective ensemble of PM/MM probe pairs. The concentration of specific and of all non-specific RNA-fragments referring to the selected probe is $c_{RNA}^S$ and $c_{RNA}^{NS}$, respectively. The second term in Eq. 7 describes progressive saturation of the probe with bound transcripts upon increasing RNA concentration according to a Langmuir isotherm.

Let us neglect saturation for sake of simplicity ($\log S_B^{PM-MM} \approx 0$). Then one obtains in the limit of high and small fractions of specific transcripts

$$\log I_B^{PM-MM,h} \approx \log K_B^{PM-MM,h} \equiv \log \frac{K_B^{PM,h}}{K_{Bc}^{MM,h}} \quad with \quad h = \begin{cases} S & for \quad c_{RNA}^S >> c_{RNA}^{NS} \\ NS & for \quad c_{RNA}^S << c_{RNA}^{NS} \end{cases}. \quad (8)$$

In other words, the middle base related log-intensity difference provides a measure of the affinity difference between complementary bases in DNA/RNA duplexes with specific and non-specific transcripts in terms of their binding constants.

**Base pairings in specific duplexes of PM and MM probes**

The sequence of a specific RNA target, $\xi_p^T$, is complementary compared with the sequence of the respective PM probe, $\xi_p^{PM}$. Consequently, the binding constant of specific hybridization of the PM, $\log K_B^{PM,S} \equiv \left\langle \log K_p^{PM}(\xi_p^{PM}\xi_p^T) \big|_B \right\rangle_{chip}$, defines the mean affinity of PM/target duplexes with the central WC pair B•b$^c$ (B= $\xi_{p,13}^{PM}$, base at position k=13 of the PM sequence) whereas the binding constant of the MM, $\log K_B^{MM,S} \equiv \left\langle \log K_p^{MM}(\xi_p^{MM}\xi_p^T) \big|_B \right\rangle_{chip}$, specifies the mean affinity of MM/target duplexes with the central SC pair B•b (B= $\xi_{p,13}^{MM}$). Figure 10 illustrates this situation for B=G.

**Fig. 10**

Let us split the middle-base related binding constant of specific hybridization into two factors according to

$$K_B^{P,S} = \kappa_B^{P,S} \cdot \kappa_{\neq 13}^{P,S} \quad , \quad (9)$$

where $\log \kappa_B^{P,S} \equiv \left\langle \log K_p^P(\xi_{p,13}^P \xi_p^T) \big|_B \right\rangle_{chip}$ is the mean effective binding constant due to the middle-base B, (P=PM, MM), and $\log \kappa_{\neq 13}^{P,S} \equiv \left\langle \log K_p^P(\xi_{p,\neq 13}^P \xi_p^T) \right\rangle_{chip}$ is the mean binding constant referring to the bases of the remaining sequence at base positions k=1-12 and 14-25 of the sequence.

The effective binding constants of the middle base of a PM/MM probe pair can be transformed into the scale of reduced Gibbs free energy of duplex formation according to

$$\begin{aligned}
-\log \kappa_B^{PM,S} &\equiv \varepsilon_{13}^{WC}(B) = \varepsilon_{0,13}^{WC} + \Delta\varepsilon_{13}^{WC}(B) \quad and \\
-\log \kappa_{Bc}^{MM,S} &\equiv \varepsilon_{13}^{SC}(\underline{B}^c) = \varepsilon_{0,13}^{SC} + \Delta\varepsilon_{13}^{SC}(\underline{B}^c)
\end{aligned} \quad . \quad (10)$$

Here $\varepsilon_{13}^{WC}(B) \equiv \varepsilon_{13}(B \bullet b^c)$ denotes the mean effective free energy (in units of the thermal energy RT) due to the formation of the WC pairs B•b$^c$ at position 13 of the probe sequence in DNA/RNA oligonucleotide duplexes on the microarray. The respective free energy of the SC pair B$^c$•b$^c$ is $\varepsilon_{13}^{SC}(B^c) \equiv \varepsilon_{13}(B^c \bullet b^c)$. We decomposed the free energies into a base-independent mean contribution, $\varepsilon_{0,13}^W \equiv \langle \varepsilon_{13}^W(B) \rangle_B$, and a base-dependent incremental contribution, $\Delta\varepsilon_{13}^W \equiv \varepsilon_{13}^W(B) - \langle \varepsilon_{13}^W(B) \rangle_B$ (W=WC, SC).

It seems safe to assume $\kappa_{\neq 13}^S \equiv \kappa_{\neq 13}^{PM,S} \approx \kappa_{\neq 13}^{MM,S}$ because the sequences of the PM and MM probes of one pair are identical except the middle base. With this approximation and making use of Eqs. 9 and 10 one obtains for the log-difference of the middle base-related binding constants of specific hybridization,

$$-\log \kappa_B^{PM-MM,S} \approx \varepsilon_{13}^{WC-SC}(B - \underline{B}^c) = \varepsilon_{0,13}^{WC-SC} + \Delta\varepsilon_{13}^{WC-SC}(B - \underline{B}^c) \quad (11)$$

*with*

$$\varepsilon_{0,13}^{WC-SC} = \varepsilon_{0,13}^{WC} - \varepsilon_{0,13}^{SC} \quad and \quad \Delta\varepsilon_{13}^{WC-SC}(B - \underline{B}^c) = \Delta\varepsilon_{13}^{WC}(B) - \Delta\varepsilon_{13}^{SC}(\underline{B}^c)$$

It consequently provides the mean free energy difference of specific duplex formation for all PM/MM probe pairs of the chip possessing PM with middle base B owing to the replacement of a SC by a WC pair, $B^c \bullet b^c \rightarrow B \bullet b^c$.

**Base pairings in non-specific duplexes of PM and MM probes**

By non-specific binding we imply the ensemble of lower affinity mismatched duplexes involving sequences other than the intended target. Hence, the effective binding constant of non-specific hybridization includes averaging over all relevant RNA fragments which only partly match the considered probes by WC pairs (see (Binder et al., 2004b) and (Binder et al., 2005)). It consequently represents the concentration-weighted average over the binding constants of a "cocktail" of RNA sequences, $\xi$, that differ from the target sequence, $\xi_T$,

$$\log K_B^{P,NS} = \left\langle \log \left\langle K_p^P(\xi_p^P \xi) \right\rangle_{\xi \neq \xi^T} \Big|_B \right\rangle_{chip} \equiv \left\langle \log \left( \sum_{\xi \neq \xi^T} c_{RNA}(\xi) \cdot K_B^P(\xi_p^P \xi) / \sum_{\xi \neq \xi^T} c_{RNA}(\xi) \right) \Big|_B \right\rangle_{chip} \quad (12)$$

Let us split the binding constant of non-specific hybridization in the effective binding constants due to the middle-base at position k=13, $\log \kappa_B^{P,NS} \equiv \left\langle \log \left\langle K_p^P(\xi_{p,13}^P \xi) \right\rangle_{\xi \neq \xi^T} \Big|_B \right\rangle_{chip}$, and due to the bases at the remaining base positions k=1-12 and 14-25 of the probe sequence, $\log \kappa_{\neq 13}^{P,NS} \equiv \left\langle \log \left\langle K_p^P(\xi_{p,\neq 13}^P \xi) \right\rangle_{\xi \neq \xi^T} \Big|_B \right\rangle_{chip}$, in analogy with the approximation used in the limiting case of specific hybridization (see previous section, Eq. 9). The effective binding constant of the middle base B is given by the weighted average over the *Boltzmann* factor of the WC and non-WC base pairings in non-specific DNA-probe/RNA dimers,

$$\kappa_B^{P,NS} \approx \sum_{b=a,u^*,g,c^*} f_{13}(\xi_{13} = b) \cdot \exp(-\varepsilon_{13}(B \bullet b)) \approx f_{13}^{WC} \cdot \exp(-\varepsilon_{13}^{WC}(B)) \quad (13)$$

Here $\varepsilon_{13}(B \bullet b)$ denotes the reduced free energy of the base pairing $B \bullet b$ (b=a,u*,g,c*). The weighting factor, $f_{13}(\xi_{13}=b)$, is the probability of occurrence of base b in $B \bullet b$ pairings.

The right-hand side of Eq. 13 assumes that only WC pairings significantly contribute to the stability of non-specific duplexes at this position. This assumption is justified, at least in a simple approach, because the interaction free energy of the strongest non-WC pair, T•g, is considerably weaker by more than 2 - 3·RT (i.e., more than 4 - 7 kJ/mole) than the free energy of the respective WC pairs, T•a and C•g (Sugimoto et al., 2000) (see also (Kierzek et al., 1999; Peyret et al., 1999; Sugimoto et al., 1995)). The stability of non-WC pairs further decreases according to T•g>>G•u≈ G•g>G•a≈A•g≈ C•a> A•a≈T•u≈C•u> A•c≈T•c (Sugimoto et al., 2000).

The logarithm of Eq. 13 shows that the binding constant in non-specific duplexes provides an effective free energy contribution which is apparently reduced by the term $\log(f_{13}^{WC})$ compared with the free energy of the WC base pairing,

$$-\log \kappa_B^{P,NS} = \varepsilon_{13}^{eff}(B) \approx \log f_{13}^{WC} + \varepsilon_{13}^{WC}(B) \qquad , \qquad (14)$$

where $f_{13}^{WC} = f(\xi_{13}=b^c)$ is the fraction of WC pairings of B in the non-specific duplexes, $0 \leq f_{13}^{WC} = N_{13}^{WC}/(N_{13}^{WC}+N_{13}^{non-WC}) \leq 1$ (the "N" denote the number of the respective pairings). Note that Eq. 14 refers to the binding of non-specific RNA fragments to P=PM and MM probes as well (see Fig. 10 for B=C). After rearrangement of Eq. 14 and making use of Eq. 10 we obtain

$$-\left(\log \kappa_B^{P,NS} + \log f_{13}^{WC}\right) = -\log \kappa_B^{PM,S} \equiv \varepsilon_{13}^{WC}(B) = \varepsilon_{0,13}^{WC} + \Delta\varepsilon_{13}^{WC}(B) \qquad , \qquad (15)$$

with $\kappa_{\neq 13}^{NS} \equiv \kappa_{\neq 13}^{PM,NS} \approx \kappa_{\neq 13}^{MM,NS}$ (see previous section) one gets for the log-difference between the binding constants of PM and MM probes in the limit of non-specific hybridization

$$-\log \kappa_B^{PM-MM,NS} \approx \varepsilon_{13}^{WC-WC}(B-B^c) = \varepsilon_{0,13}^{WC-WC} + \Delta\varepsilon_{13}^{WC-WC}(B-B^c) \qquad (16)$$

*with*

$$\varepsilon_{0,13}^{WC-WC} = \varepsilon_{0,13}^{WC}\Big|_{PM} - \varepsilon_{0,13}^{WC}\Big|_{MM} \quad and \quad \Delta\varepsilon_{13}^{WC-WC}(B-B^c) = \Delta\varepsilon_{13}^{WC}(B) - \Delta\varepsilon_{13}^{WC}(B^c)$$

Here $\varepsilon^{WC-WC}(B-B^c)$ denotes the mean free energy difference between DNA/RNA oligonucleotide duplexes with the WC pairs B•b$^c$ and B$^c$•b at position k=13 of the 25-meric DNA probe, which is averaged over all PM/MM probe pairs of the chip. The middle-base related log-difference of the binding constants of the PM and MM for non-specific hybridization consequently describes the change of free energy upon the reversal the WC pair, B•b$^c$ → B$^c$•b (see Fig. 10 for illustration).

**Fig. 11**

### The mean free energy difference between WC and SC pairings

The PM-MM differences of the log-intensity data, $\log I_B^{PM-MM}$, and the derived sensitivities, $Y_B^{PM-MM}$ and $\sigma_{13}^{PM-MM}(B)$, are directly related to the free energy of base pairings due to DNA/RNA duplex formation on the microarray. Figure 11 illustrates the base specific free energy contributions and the respective differences together with the relevant experimental intensity and sensitivity data in terms of an energy level diagram. The panel below (part a) shows the differences between the effective free energy of complementary middle bases in DNA oligonucleotide probes upon duplexe formation with non-specific (left part) and specific (right part) RNA fragments. The respective values of $\varepsilon_{13}^{WC-WC}(B•B^c)$ and $\varepsilon_{13}^{WC-SC}(B•B^c)$ were estimated by means of the log-intensity difference between PM and MM probes (see Eqs. 1, 3, 11 and 16 and also the panel above in Figs. 6 and 7).

For equally hybridized PM and MM one expects a fraction of bright MM of f(PM<MM)≈0.5 and a middle-base related mean PM-MM log-intensity difference of $\log I_B^{PM-MM} \approx 0$, in contrast to the results. Note that the middle-base related mean PM-MM log-intensity difference, $\log I_B^{PM-MM}$, and the respective fraction of bright MM, $f_B(MM<PM)$, asymmetrically distribute about the expected values at vanishing amounts of specific transcripts (see Figs. 7 and 8). As a consequence, the mean free energy difference between WC pairings in PM and MM probes, $-\varepsilon_{0,13}^{WC-WC} \approx 0.05-0.1$, significantly deviates from zero (see panel below in Fig. 3 and dashed lines in Fig. 11 for illustration). One expects however for $\varepsilon_0^{WC-WC}$ a vanishing value (see Eq. 16) because the PM and MM on the average possess an equal affinity for WC pairings with the non-specific RNA fragments. The non-random probability distribution of the middle base among the PM probes on the HG-U133 Affymetrix© chip with a slightly higher fraction of C and T (23% and 31%, respectively) compared with G and A (22% and 24%) partly, but not fully explains the significant deviation of the observed from the expected value. Possibly also a non-random base distribution of the PM and MM probes at $k \neq 13$ and of the relevant non-specific RNA fragments give rise to the observed effect because it potentially introduces an asymmetric relation between the PM and MM intensities.

The mean free energy difference considerably changes to $-\varepsilon_{0,13}^{WC-SC} \approx 0.8-0.9$ in the limit of specific hybridization (see panel below in Fig. 3 and dashed lines in part a of Fig. 11 for illustration). It provides the mean free energy difference between a WC and a SC pair in RNA-target/DNA-probe duplexes on the microarray. Interestingly the obtained value of $\varepsilon_{0,13}^{WC-SC}$ well agrees with the mean reduced free energy of a WC pair in RNA/DNA oligonucleotide duplexes in solution, $-\varepsilon_{0,sol}^{WC} = 0.75$-$1.03$, which was estimated by means of

$$\varepsilon_{0,sol}^{WC} = \left\langle \varepsilon_{sol}^{WC}(B) \right\rangle_B \quad with$$
$$\varepsilon_{sol}^{WC}(B) = (8RT \cdot \ln 10)^{-1} \sum_{X=A,T,G,C} (G(B,X) + G(X,B)) \qquad (17)$$

using the respective nearest neighbor free energy terms, G(BB'); B, B'=A,T,G,C (Sugimoto et al., 1995; Wu et al., 2002). The agreement between $\varepsilon_{0,13}^{WC-SC}$ and $\varepsilon_{0,sol}^{WC}$ can be rationalized if the strengths of base pair interactions are similar in RNA/DNA oligonucleotide duplexes in solution and on microarrays and if the mean free energy of the SC pairs is much weaker than that of the WC pairs, $|\varepsilon_{0,13}^{SC}| << |\varepsilon_{0,13}^{WC}|$.

**Base-specific interactions: the purine/pyrimidine asymmetry**

Part b of Fig. 11 illustrates the middle base-specific incremental contribution to the free energy differences between complementary bases in WC pairs (left part), and complementary bases in WC and SC pairings (right part), which were extracted from the middle base related PM-MM sensitivity difference and the single base model (see Eqs. 5 and 6 and Figs. 8 and 9). The duplet-like relation between the $\Delta\varepsilon_{13}^{WC-WC}(B)$-values in the limit of dominating non-specific hybridization can be explained by the formation of WC pairings between the middle base of the probes with the bound

RNA fragments and an asymmetry of base pair interactions upon reversal of the type B•b$^c$ → B$^c$•b as illustrated in part c of Fig. 11 (left panel, see also Fig. 10 for B=G). The common binding strength for the same base in PM and MM probes and the fact that a pyrimidine (Y=C,T) in the DNA-probe forms a stronger WC pair (C•g and T•a) than the complementary purine (R=G,A, G•c*, A•u*) give rise to the duplet-like relation $-\Delta\varepsilon_{13}^{WC-WC}(Y) = \Delta\varepsilon_{13}^{WC-WC}(R) > 0$ as indicated by the sensitivity differences $Y_B^{PM-MM,NS}$.

The duplet transforms into a triplet-like pattern of the incremental contributions, $\Delta\varepsilon_{13}^{WC-SC}(B)$ in the limit of high specific transcript concentration (see right panel of part b of Fig. 11 and Fig. 8). This relation between the sensitivities can be rationalized if the middle base of the PM probe forms a WC pair whereas the complementary middle base of the MM probe "faces itself" in a self complementary (SC) base pair with the RNA target (see Fig. 10 for illustration). The triplet-like relation between the data is compatible with the assumption that the SC pairs on the average only weakly contribute to duplex stability as stated above, i.e. $\left|\Delta\varepsilon_{13}^{SC}(B)\right| << \left|\Delta\varepsilon_{13}^{WC}(B)\right|$, and with the pyrimidine-purine asymmetry of WC pairings, $\Delta\varepsilon^{WC}(C) \approx -\Delta\varepsilon_{13}^{WC}(A) < -\Delta\varepsilon_{13}^{WC}(G) \approx \Delta\varepsilon_{13}^{WC}(T)$ (see left panel in part c of Fig. 11). In this case the different base pairings, namely the WC pair B•b$^c$ of the PM and the SC pair B$^c$•b$^c$ of the respective MM give rise to $\Delta\varepsilon_{13}^{WC-SC}(B-B^c) \approx \Delta\varepsilon_{13}^{WC}(B)$. In other words, the free energy change upon the replacement of a SC by a WC pairing, e.g., C•c* → G•c*, roughly reflects the strength of the WC pair (see Fig. 10). The respective PM-MM sensitivity differences consequently order according to the strengths of the WC pairings in DNA/RNA oligo duplexes, C > G ≈ T > A.

Note that the reduced Gibbs free energy of base pairings in DNA/RNA oligonucleotide duplexes in solution, $\Delta\varepsilon_{sol}^{WC}(B) = \varepsilon_{0,sol}^{WC} - \left\langle\varepsilon_{sol}^{WC}(B)\right\rangle_B$ (see Eq. 17), decreases in a similar order according to C > G > T > A. Hence, the base pair interactions derived from solution data also show a purine/pyrimidine asymmetry. It can be specified by the asymmetry parameter, which characterizes the relative gain of free energy upon the reversal of the bond direction according to R•y→Y•r, $A_{sol}^{WC}(C•g/G•c) \equiv -\{\Delta\varepsilon_{sol}^{WC}(C•g) - \Delta\varepsilon_{sol}^{WC}(G•c)\}/\{|\Delta\varepsilon_{sol}^{WC}(C•g)| + |\Delta\varepsilon_{sol}^{WC}(G•c)|\} \approx 0.3\pm0.1$ and $A_{sol}^{WC}(T•a/A•u) \approx 0.4\pm0.1$. The respective asymmetry increases to $A_{13}^{WC}(C•g/G•c^*) \approx A_{13}^{WC}(T•a/A•u^*) \approx 0.9\pm0.1$ for the pairings of the middle base of microarrays oligo probes (Binder et al., 2004a). Note that the WC base pairings of the purines on the microarray, G•c* and A•u*, carry the biotinyl and the fluorescent label. Hence, the higher purine/pyrimidine asymmetry on the microarray can be attributed to the labelling of the RNA fragments, which potentially hampers binding (Binder et al., 2004a; Naef and Magnasco, 2003).

**The PM/MM asymmetry of probe intensities**

Our interpretation of non-specific hybridization on microarrays assumes that the hybridization solution contains a sufficient large number of different sequences, which partially match the probe sequences via WC pairings including their central bases. In other words, this "cocktail" of RNA fragments with a broad distribution of base compositions on the average enables WC pairings with the middle bases of the PM and with the complementary middle base of the respective MM as well. As a consequence, the base-related affinities are virtually equal for base B in both types of probes but different for the complementary couples of bases B and $B^c$ of each PM/MM probe pair.

This asymmetric relation of base-pair interactions in non-specific duplexes gives rise to observed asymmetry of probe intensities, i.e., the tendency of "bright" PM for B=C,T, and, vice versa, of "bright" MM for B=G,A. The "riddle of bright MM" refers solely to non-specific hybridization. It simply reflects the reversal of WC pairings with asymmetrical binding strength according to our interpretation. The results of previous analyses of the PM-MM intensity relation of all probe pairs of a series of GeneChips (Naef et al., 2002; Naef and Magnasco, 2003) can be understood if the overwhelming majority of the probes of the chips are non-specifically hybridized.

In the special case of specific hybridization each probe is related to only one specific RNA-target sequence, which completely matches the sequence of the PM probe via WC pairings. The complementary middle base of the MM consequently mismatches the respective position of the target sequence via a SC pairing. Our analysis reveals that almost no of the analyzed 462 spiked-in probe pairs gives rise to "bright MM" if specific transcripts dominate hybridization. This result strongly indicates a considerably reduced affinity of the mismatch, which causes the significantly reduced intensity of the MM compared with that of the PM.

Using a stochastic approach, Wu and Irizzary (Wu and Irizarry, 2004) claimed that the effect of bright MM is a consequence of the noisy character of the system and of the difference in the affinities for different sequences combined with the assumption that the MM do not measure specific signal. Our results however clearly indicate that also the MM bind specific transcripts in relevant amounts. Moreover, the analysis of chip data without differentiation between specific and non-specific hybridization seems not appropriate at least at small intensities because the central base affects duplex formation in a letter-specific fashion.

**Accuracy and precision of expression measures**

The basic application of the GeneChip technology intends to estimate the level of differential gene expression in terms of the log-fold change of the RNA transcript concentration between different samples, $DE^{true} \equiv \log\{c_{RNA}^S(samp)/c_{RNA}^S(ref)\}$, for example, between the sample of interest and an appropriately chosen reference. The respective log-intensity ratio, $DE_B^P \equiv \log\{I_B^P(samp)/I_B^P(ref)\}$ with P=PM,MM, provides a measure of the differential expression in the simplest approach. In the Appendix we show that $DE_B^P$, the apparent differential expression, additively decomposes into the true log-fold change of the RNA concentration and an incremental contribution $\Delta DE_B^P$,

$$DE_B^P = DE^{true} + \Delta DE_B^P \quad with \quad \Delta DE_B^P \equiv \log\frac{1+r_c(samp)\cdot r_B^P}{1+r_c(ref)\cdot r_B^P} \tag{18}$$

**Fig. 12**

The latter term is a function of the concentration ratio of non-specific and specific RNA, $r_c \equiv c_{RNA}^{NS}/c_{RNA}^{S}$ in the reference and the sample, and of the ratio of the respective binding constants, $r_B^P \equiv K_B^{P,NS}/K_B^{P,S}$. It specifies the deviation of the apparent differential expression from its true value and thus it characterizes the accuracy of the estimated $DE_B^P$-value. Figure 12 (panels a and b) shows $DE_B^P$ for P=PM,MM as a function of $DE^{true}$ using the interaction parameters determined in this study (see the Appendix for details). The apparent values systematically underestimate the differential expression owing to the non-specific background intensity not related the concentration of the target RNA. Note that the MM-only estimates are less accurate compared with the PM-only values, i.e., $|\Delta DE^{MM}_B| > |\Delta DE^{PM}_B|$, because the non-specific background provides a larger contribution to the MM intensity on a relative scale.

The MM probes were designed to estimate the amount of non-specific hybridization, and, this way, to provide corrected intensities by means of the intensity difference of the probe pairs, $\Delta \equiv PM-MM$ (see Appendix). Indeed, the respective differential expression values on average provide a relative accurate result (see Fig. 12, part c). The averages of the $DE_B^P$ over the four middle bases show that the accuracy of the intensity measures of the differential expression decrease according to "true" $\approx$ PM-MM > PM > MM (see Fig. 12, part d).

Interestingly, the calculated $DE_B^P$-data reveal a second effect. The PM-only estimates, $DE_B^{PM}$, are independent of the middle base whereas the log-fold intensity changes of the MM and consequently also that of the PM-MM difference markedly vary as a function of B=A,T,G,C. This effect can be rationalized by the fact that the specific and non-specific duplexes of the PM are both characterized by the same WC pairing in the middle of the sequence whereas the MM form a SC pair in the specific duplexes and a WC pair in the non-specific ones (see Fig.10). Consequently, the interaction- and consequently also the intensity-characteristics vary in a similar fashion for all middle bases in the PM duplexes upon changing the concentration ratio $r_c$ whereas the respective interactions in the MM duplexes vary differently.

The middle base of the probes consequently introduces a systematic source of variability to the apparent differential expression values, $DE_B^P$, because microarray probes are usually designed without special attention to their middle base. Panel e of Fig. 12 shows the coefficient of variation of the apparent log-fold changes, $CV(DE^P_B) \equiv SD(DE^P_B)/<DE^P_B>$ (SD and <...> denote the standard deviation and the arithmetic average, respectively), as a measure of the variability upon changing B. It is inversely related to the precision (resolution) of the respective differential expression measures. The precision of the PM-only intensity measure clearly outperforms those of the two other estimates, i.e. PM > MM $\approx$ PM-MM.

Hence, the high accuracy of expression measures based on the PM-MM intensity difference is opposed by their relatively low precision. The latter effect depends in a systematic fashion on the middle base. Its explicit consideration and correction in sophisticated analysis algorithms which take into account the middle base specific intensity characteristics is expected to improve the precision of PM-MM measures.

**Hybridization on microarrays**

Melting experiments on DNA oligonucleotide hybridization on microarrays have shown that surface tethered DNA duplexes are less stable than hybrids formed in bulk solution as indicated by the substantial reduction of the standard enthalpy change upon denaturation (Watterson et al., 2000). These results suggest that the physical environment of hybrids formed at the solid interface is significantly different from that in solution owing to kinetic effects (Chan et al., 1995), equilibrium thermodynamics (Bhanot et al., 2003) and surface electrostatics (Chan et al., 1995; Vainrub and Pettitt, 2002). The latter effect causes, e.g., the Coulomb blockage of microarray hybridization with increasing coverage of the array probes (Halperin et al., 2004; Peterson et al., 2001; Vainrub and Pettitt, 2002).

On the other hand, the thermodynamic parameters of surface hybridization and thus the stability of the hybrids on microarrays display the same general trends with respect to changes of solution ionic strength and the presence of single mismatches as the duplexes formed in bulk solution (Watterson et al., 2000). These results agree with our recent findings, which indicate agreement between chip and solution data with respect to the specificity of base pair interactions on one hand side and differences between both systems with respect to the absolute magnitude of the interactions strength on the other hand (Binder et al., 2004a). In particular we found that the base-specific nearest neighbour free energies of WC base pairings in DNA/RNA duplexes on microarrays strongly correlate with that for hybridization in solution whereas their magnitude is considerably decreased compared with the solution data.

Surface hybridization is obviously well compatible with hybridization in solution with respect to the relative stability of base pairings. The present study confirms this "conventional" view on microarray hybridization. It predicts that (i) non-specific binding is on the average identical for PM and MM probes with systematic deviations owing to the pyrimidine/purine asymmetry of WC base pair interactions in RNA/DNA duplexes, and that (ii) the mismatch reduces the affinity of specific target-binding to the MM due to the considerably weaker interactions of mismatched base pairings.

In this study we used two independent measures to estimate duplex stability as a function the middle base, namely the positional dependent SB-sensitivities and the sensitivity-averages over probes with a common middle base. This simple description in terms of single-base related parameters to a large extent neglects cooperative effects of the whole sequence of the oligonucleotides. The explicit consideration of the adjacent bases in terms of nearest neighbor- and/or middle triple-related energy parameters is expected to refine the results (Binder et al., 2004a). Moreover, also the propensity of the probe and of the target for intramolecular folding (Matveeva et al., 2003), "zippering effects" (i.e., target/probe duplexes which look like a partly opened double-ended zipper (Deutsch et al., 2004)) and a certain fraction of shorter oligonucleotide lengths after imperfect photolithographic synthesis (Jobs et al., 2002; McGall et al., 1997) modifies the duplex stability with possible consequences for the middle base-related interaction parameters.

Note that the positional dependent SB-sensitivity terms are effective parameters, which are averaged over all possible microscopic states of the respective duplexes. The contribution of each base pairing is weighted by its probability to occur in the individual DNA/RNA dimers. Consequently "zippering effects" and/or shorter probe lengths can explain the observed sensitivity gradient along the sequence (see panel above in Fig. 9) because the probability of paired bases decreases in direction towards the ends in the zippered and/or truncated duplexes. On the other hand, these effects are minimum in the centre of the sequence and, moreover, they affect the paired PM and MM in a similar fashion leaving the PM/MM log-intensity difference, and thus the estimated middle base related affinity parameters virtually unaffected.

**Summary and Conclusions**

Specific and non-specific hybridization give rise to different relations between the PM and MM intensities, namely a triplet-like pattern of the PM-MM log-intensity difference in the former case and a duplet-like split in the latter case. The analysis of intensity data without the careful separation between specific and non-specific binding events can therefore lead to confusion about "what RNA hybridizes the probes" and in consequence to the incorrect assignment of base pair interactions. This in turn affects the estimation of signal intensities in terms of gene expression and, in particular, the consideration of the MM intensities as a correction term for non-specific hybridization of the PM.

It has been shown that relevant interaction parameters for estimating probe intensities can be derived from chip data, and, in particular, that the set-averaged probe intensity as a simple intensity-criterion allows to discriminate between predominantly specifically and predominantly non-specifically hybridized probes. Here we analyzed the PM and MM intensities in terms of simple single base-related parameters to establish the basic relations between the PM and MM data. A more detailed approach using nearest-neighbor interaction parameters is expected to refine the results.

The analysis indicates that the intensity of complementary MM introduces a systematic source of variation compared with the intensity of the respective PM probe. In consequence, the naive correction of the PM signal by subtracting the MM intensity decreases the precision of expression measures. Our results suggest improved algorithms of data analysis, which explicitly consider the middle-base related bias of the MM intensities to reduce their systematic effect. Moreover, the knowledge of the central base pairings in specific and non-specific duplexes allows revision of mismatch-based strategies of chip design, for example, by testing alternative rules for predefined mismatches than the complementary mismatches used on GeneChips.

**Acknowledgments**

We thank Prof. Markus Loeffler and Prof. Peter Stadler for support and discussion of aspects of the paper. The work was supported by the Deutsche Forschungsgemeinschaft under grant no. BIZ 6-1/2.

**Appendix: Derivation of Eq. 18**

The middle base averaged probe intensity can be approximated by the superposition of contributions due to specific and non-specific hybridization, $I_B^P = I_B^{P,S} + I_B^{P,NS}$, if one neglects saturation for sake of simplicity. The intensities of the specifically and non-specifically hybridized probes are directly related to the concentrations and the binding constants of the respective RNA fragments, i.e., $I_B^{P,h} \approx F \cdot c_{RNA}^h \cdot K_B^{P,h}$ (h=S,NS; F is a constant). With Eq. 9 one obtains after some rearrangements

$$I_B^P \approx F \cdot \kappa_{\neq 13}^S \cdot \kappa_B^{P,S} \cdot c_{RNA}^S \cdot \left(1 + r_c \cdot r_B^P\right)$$

$$\text{with} \quad r_c = \frac{c_{RNA}^{NS}}{c_{RNA}^S} \quad \text{and} \quad r_B^P \equiv \frac{K_B^{P,NS}}{K_B^{P,S}} \approx \left(f^{WC}\right)^{N_b - 1} \cdot \frac{\kappa_B^{P,NS}}{\kappa_B^{P,S}} \tag{A1}$$

The latter equation assumes $\kappa_{\neq 13}^{NS} = \left(f^{WC}\right)^{N_b - 1} \cdot \kappa_{\neq 13}^S$, i.e. a constant and positional independent fraction of WC pairings of $f^{WC} \approx f_{13}^{WC}$ for each of the $N_b=25$ sequence position in the non-specific duplexes in analogy with Eq. 13. The ratio of the binding constants can be further specified using Eq. 15,

$$r_B^P = \left(f^{WC}\right)^{N_b} \cdot \begin{cases} 1 & \text{for} \quad P = PM \\ \kappa_B^{PM,S} / \kappa_B^{MM,S} & \text{for} \quad P = MM \end{cases} \tag{A2}$$

$$\text{with} \quad \kappa_B^{PM,S} / \kappa_B^{MM,S} = \exp\left\{-\ln 10 \cdot (\varepsilon_{0,13}^{WC-SC} + \Delta\varepsilon_{13}^{WC}(B) - \Delta\varepsilon_{13}^{SC}(B))\right\}$$

Analogous considerations lead to the result that Eq. A1 applies also to the intensity difference between PM and MM probes, $I_B^\Delta \equiv I_B^{PM} - I_B^{MM}$, with the substitutions for P=Δ

$$\kappa_B^{\Delta,h} = \kappa_B^{PM,h} \cdot \left(1 - E_{Bc,B}^h\right) \quad \text{with} \quad E_{Bc,B}^h \equiv \kappa_{Bc}^{MM,h} / \kappa_B^{PM,h} \quad ; \quad h = S, NS \quad \text{and}$$

$$E_{Bc,B}^S \approx \exp\left\{\ln 10 \cdot (\varepsilon_{0,13}^{WC-SC} + \Delta\varepsilon_{13}^{WC-SC}(B - B^c))\right\} \tag{A3}$$

$$E_{Bc,B}^{NS} \approx \exp\left\{\ln 10 \cdot (\varepsilon_{0,13}^{WC-WC} + \Delta\varepsilon_{13}^{WC-WC}(B - B^c))\right\}$$

Equation 18 can be directly obtained by application of Eq. A1 for two transcript concentrations, a "sample" and the "reference", and its insertion into $DE_B^P \equiv \log\{I_B^P(\text{samp})/I_B^P(\text{ref})\}$ with P=PM,MM,Δ. The incremental contribution, $\Delta DE_B^P \equiv \log \frac{1 + r_c(\text{samp}) \cdot r_B^P}{1 + r_c(\text{ref}) \cdot r_B^P}$, was estimated using Eqs. (A2) and (A3) and the following parameters obtained in this study: $\varepsilon_{0,13}^{WC-WC} \approx -0.05$; $\varepsilon_{0,13}^{WC-SC} \approx -0.85$; $\Delta\varepsilon_{13}^{WC}(B) \approx 0.25, 0.05, -0.05, -0.25$ (B=A,T,G,C) and $\Delta\varepsilon_{13}^{SC}(B) \approx -0.05, 0.0, 0.0, 0.05$. The factor $(f^{WC})^{N_b} \approx 10^{-2.5}$ was estimated previously (Binder et al., 2005). The spiked-in experiment used a common concentration level of non-specific RNA fragments ($c^{NS}(\text{samp}) \approx c^{NS}(\text{ref})$), which gives rise to the following relation between the concentration ratios of the sample and the reference:

$$r^c(samp) = \frac{c^{NS}(samp)}{c^S(samp)} = r^c(ref) \cdot 10^{-DE^{true}}.$$

**Figure Captions**

**Figure 1**: Log-intensity difference, $\log I^{PM-MM} = \log I^{PM} - \log I^{MM}$, of the spiked-in probes taken from the LS experiment as a function of the mean set averaged intensity, $\langle \log I^{PM+MM} \rangle_{set} = 0.5 \langle (\log I^{PM} + \log I^{MM}) \rangle_{set}$, which serves as an approximate measure of the specific transcript concentration. Intensity averages over the probe sets are shown by open circles. The panel below shows the log-differences for three selected spiked-in concentrations. Each concentration spans a range of about $\delta \langle \log I^{PM+MM} \rangle \approx \pm 0.5$ as indicated by the lines between the two panels. Note that the log-intensity-difference shifts upwards with increasing $\langle \log I^{PM+MM} \rangle_{set}$ indicating the progressive decrease of the fraction of "bright" MM with increasing amount of specific transcripts.

**Figure 2**: The fraction of bright MM, f(MM<PM) (panel below) and the mean log-intensity difference, $\langle \log I^{PM-MM} \rangle_{sp-in}$ (panel above), of the spiked-in probes taken from the LS experiment strongly correlate with the concentration of specific transcripts. The respective fraction of probe sets, $f^{set}$(MM<PM), meeting the condition $\langle \log I^{PM-MM} \rangle_{set} < 0$ is shown by triangles in the panel below. The data can be well explained by the probability that more than n(min)=6 – 7 individual probe pairs of the set independently possesses bright MM using the Binominal distribution (see lines denoted by "6" and "7", respectively).

**Figure 3**: Log-intensity difference between PM and MM probes of the whole data set of about 250.000 probes of a HG U133 chip (panel above), fraction of bright MM (panel below, left ordinate) and mean log-intensity difference (panel below, right ordinate) as a function of the mean set averaged intensity. The fraction of bright MM and the mean difference were calculated as running averages over 1000 subsequent probes along the abscissa. Note the agreement with the respective data obtained from the spiked-in data set (Figs. 1 and 2). It shows that the dependence of the probe intensities on the concentration of specific transcripts applies to the whole set of probes of the chip.

**Figure 4**: The figure shows the same type of data as in Fig. 3, however only probe pairs with a G and C in the middle of the PM sequence are selected (see Figure for assignments). The data referring to the pyrimidine and purine middle base are shifted in vertical direction to each other. Compare with Fig. 5 and see also legend of Fig. 3.

**Figure 5** The figure shows the same type of data as in Fig. 3, however only probe pairs with a T and an A in the middle of the PM sequence are selected (see Figure for assignments). Compare with Fig. 4 and see also legend of Fig. 3.

**Figure 6**: Fraction of bright MM (panel below) and mean log-intensity difference (panel above) for probe pairs with a B= A,T,G,C in the middle of the PM sequence (see Figure for assignments) as a function of the mean set averaged intensity. The data were replotted from Figs. 4 and 5 (see the respective legends for details). The data refer to the whole data set of about 250.000 probes of a HG U133 chip. Note that the log-intensity differences split in to a duplet-like pattern at small abscissa values referring to non-specific hybridization and into a triplet-like pattern at high abscissa values referring to specific hybridization (see part above).

**Figure 7**: Fraction of bright MM (panel below) and mean log-intensity difference (panel above) for probe pairs with a B= A,T,G,C in the middle of the PM sequence (see Figure for assignments) as a function of the concentration of specific transcripts. The data refer to the spiked in data set of 462 different probes. Compare with Fig. 6. Both Figures show essential identical properties for the spiked-in and the full set of probes.

**Figure 8**: Middle base related sensitivity of probe pairs with a B= A,T,G,C in the middle of the PM sequence (see Figure for assignments and Eq. 2) as a function of the concentration of specific transcripts. The concentration ranges of dominating non-specific (NS) and of specific (S) hybridization are indicated by vertical dotted lines. The duplet in the limit of non-specific hybridization transforms into a triplet like pattern in the limit of specific hybridization. The sensitivity provides a measure of the base specific contribution to the free energy of RNA/DNA duplex stability.

**Figure 9**: Positional dependent single base sensitivity profile of the PM (symbols) and MM (lines) probes in the limit of non-specific (left part) and specific (right part) hybridization. The two panels below show the respective PM-MM difference profiles (see Eq. 5). Note that the PM-MM difference of the middle base considerably exceeds the contributions of the bases at the remaining positions along the sequence.

**Figure 10**: Schematic illustration of the base pairing in the middle of the sequence of PM (left) and MM (right) probes upon duplex formation with specific (above) and non-specific (below) transcripts. The example shows a probe pair with middle bases G and C of the PM and MM probes, respectively. Upper case letters refer to the DNA probes and lower case letters to the RNA transcripts (the asterisk indicates labeling). The middle base effectively forms Watson Crick pairings in the non-specific duplexes of the PM as well in the non-specific duplexes of the MM (i.e., C•g and G•c* in the chosen example, respectively). It also forms a Watson Crick pair in the specific duplexes of the PM probes but a self complementary pair in the specific duplexes of the MM probes (i.e., C•g for the PM and G•g for the MM.). Note that the remaining positions along the probe sequences are partly mismatched in the non-specific duplexes.

**Figure 11**: Schematic energy level diagram of the Gibbs free energy of base pairings and their differences at the central position of PM and MM probes in the limit of non-specific (left part) and specific (right part) hybridization. Part a: difference of the respective total free energy contribution of complementary bases (see Eqs. 11 and 16); part b: difference of the base-specific incremental contribution; part c: base-specific incremental free energy contribution. The free energy terms were estimated using the log-intensity difference, $\log I_B^{PM-MM}$ (part a, compare with Figs. 3-5), the sensitivity differences $Y_B^{PM-MM}$ and $\sigma_{13}^{PM-MM}(B)$ (part b, compare with Fig. 8) and the single base sensitivity terms, $\sigma_{13}^{PM}(B)$ and $\sigma_{13}^{MM}(B)$ (part c). See text.

**Figure 12**: Apparent differential expression, $DE_B^P$, as a function of the "true" log-fold change of the RNA-target concentration, $DE^{true}$. The apparent values were calculated using the log-fold change of the probe intensities as described in the Appendix (see also Eq. 18). The PM-only (panel a) and MM-only (panel b) intensity data underestimate the true value whereas the PM-MM intensity difference provides an acceptable measure of $DE^{true}$ (panel c). Note that $DE_B^P$ depends on the middle base B=A,T,G,C for P=MM and PM-MM. Panels d and e show the mean values, $<DE_B^P>$, which are averaged over the four possible middle bases and the respective coefficient of variation, $CV(DE_B^P)$, respectively. The deviation of $<DE_B^P>$ from $DE^{true}$ specifies the accuracy and $CV(DE_B^P)$ is inversely related to the precision of the respective measure of gene expression (see text).

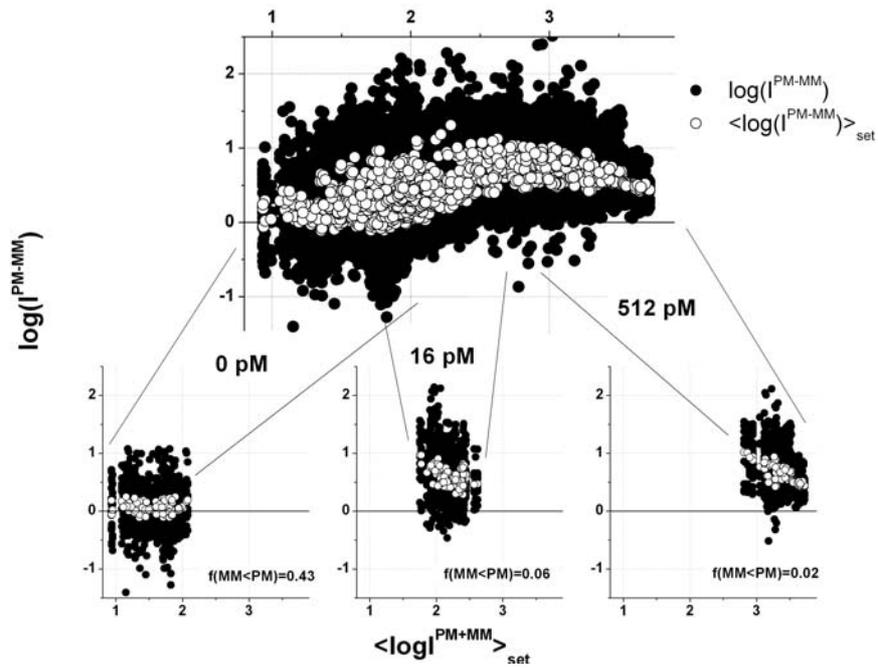

**Figure 1, Binder and Preibisch**

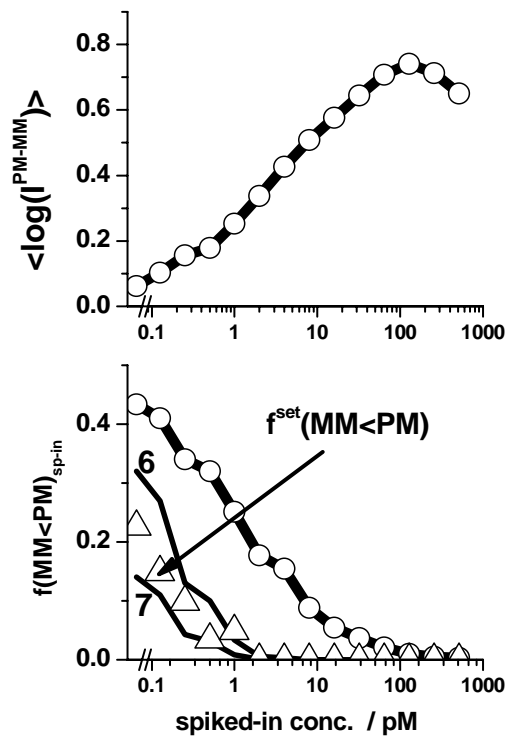

**Figure 2, Binder and Preibisch**

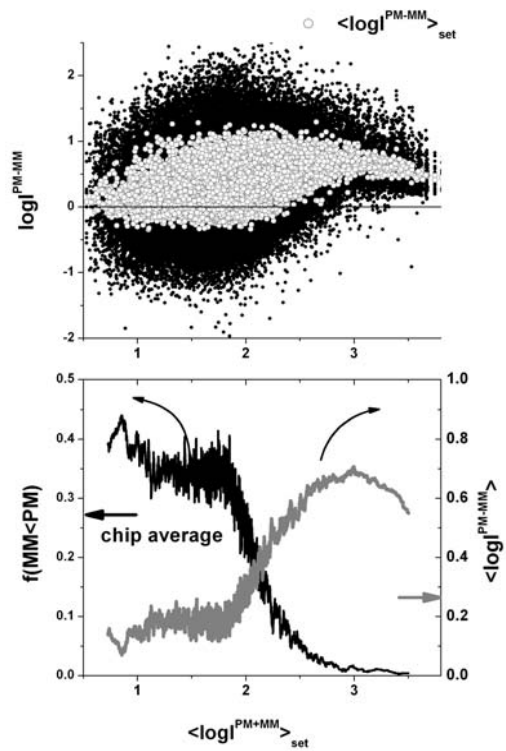

**Figure 3, Binder and Preibisch**

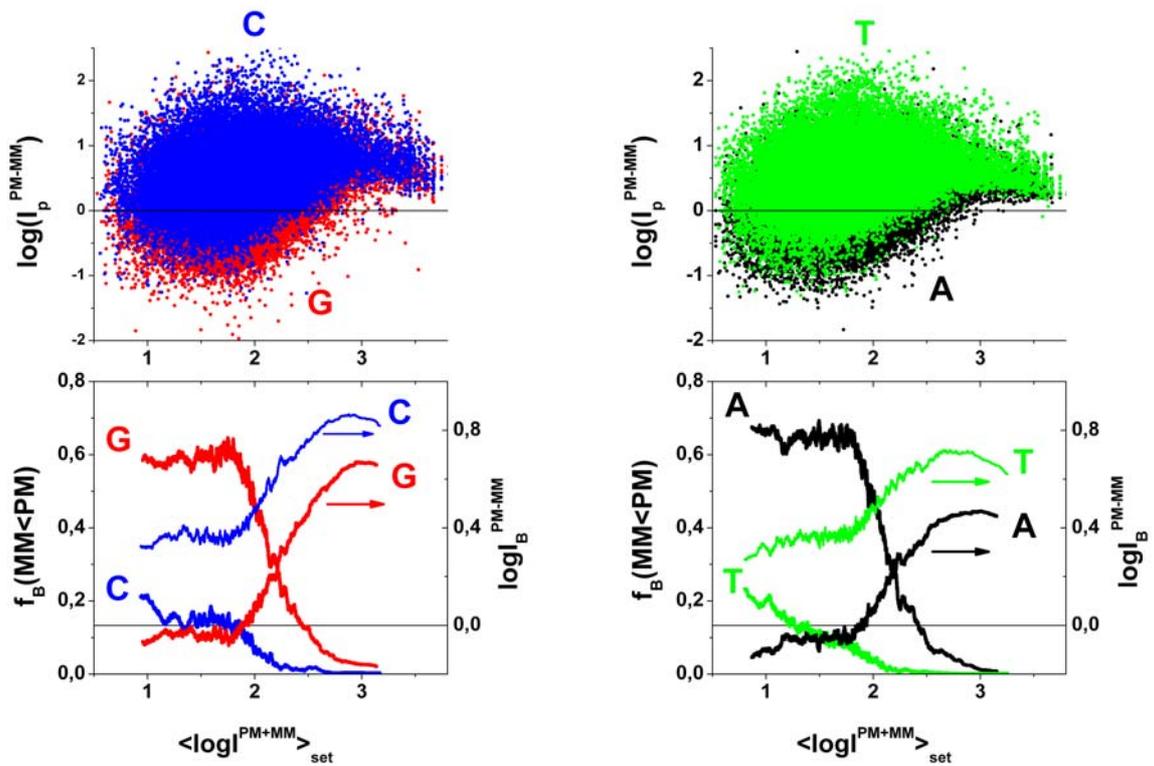

**Figure 4, Binder and Preibisch (left)**
**Figure 5, Binder and Preibisch (right)**

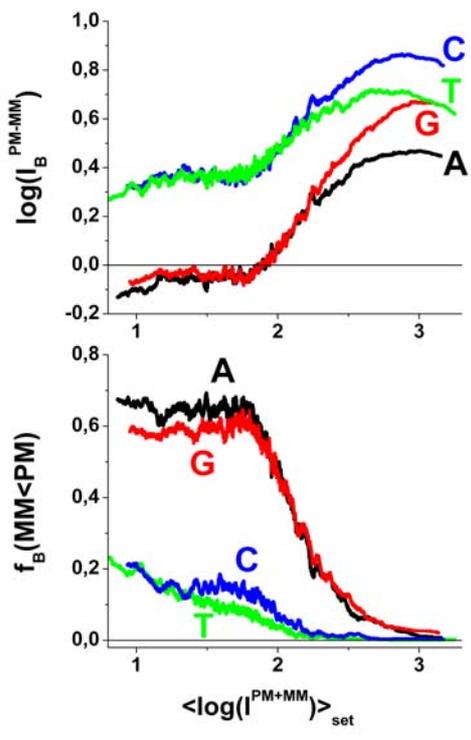 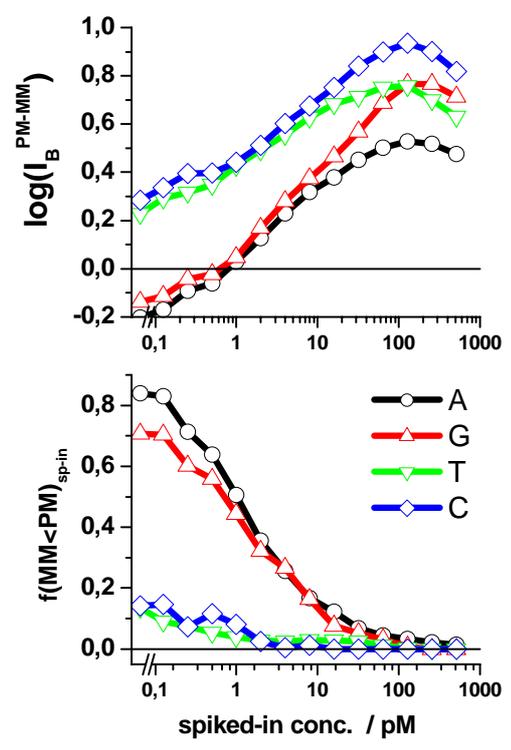

**Figure 6, Binder and Preibisch (left)**

**Figure 7, Binder and Preibisch (right)**

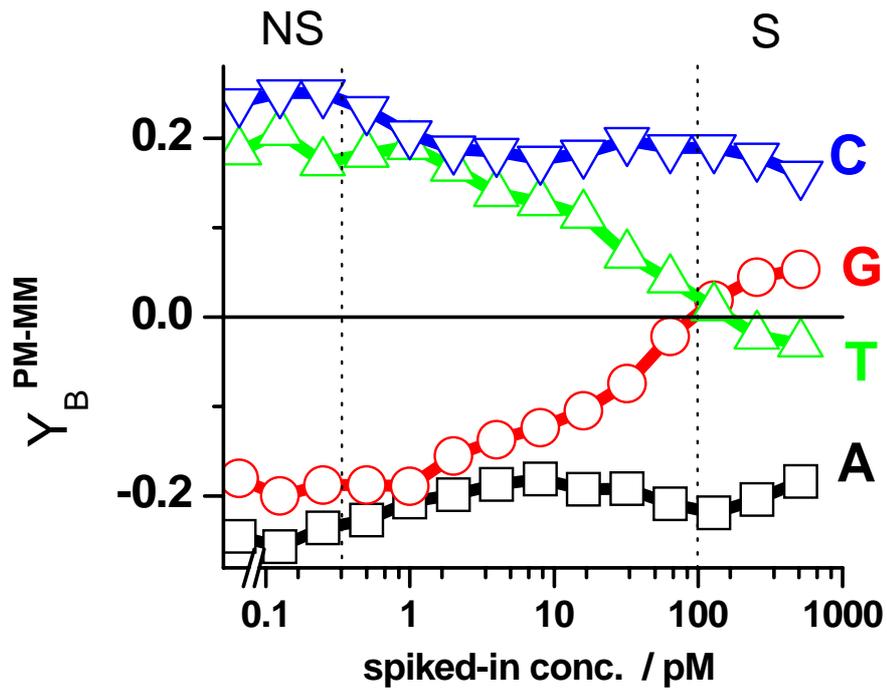

**Figure 8, Binder and Preibisch**

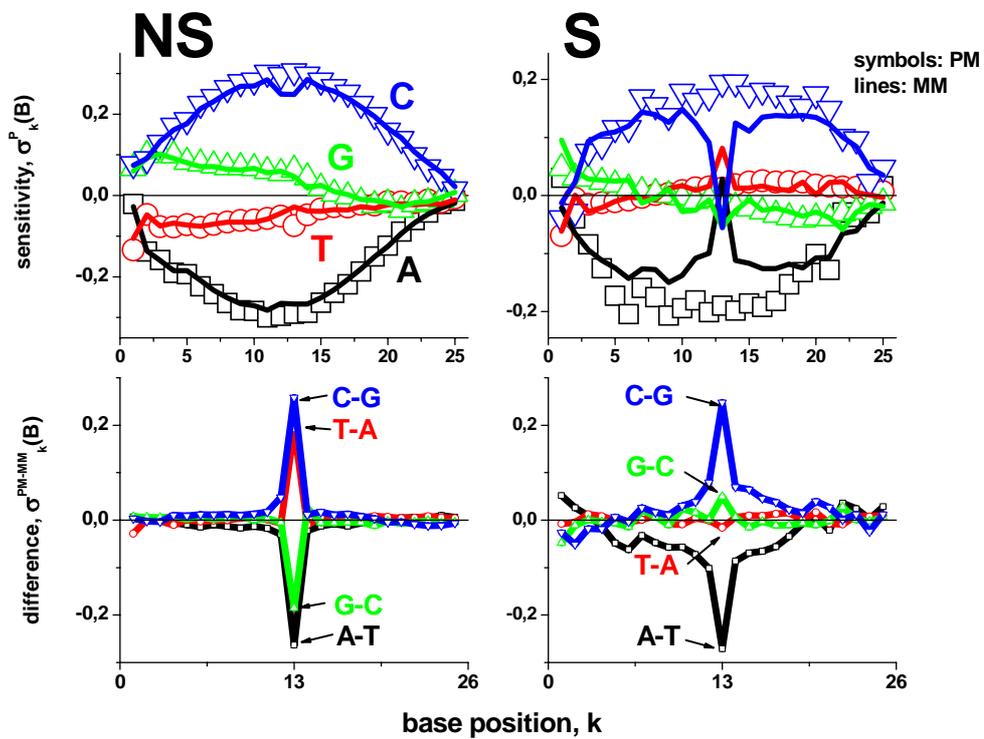

**Figure 9, Binder and Preibisch**

**specific hybridisation:**

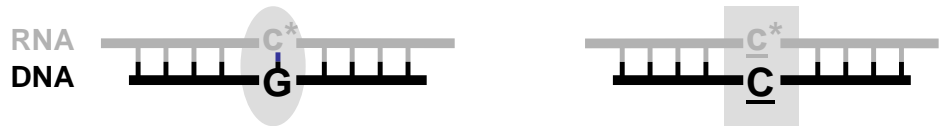

**non-specific hybridisation:**

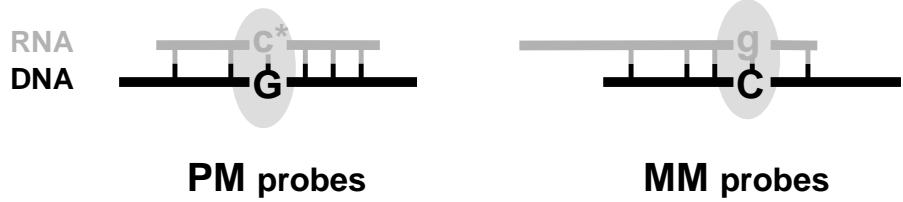

**PM** probes          **MM** probes

**Figure 10, Binder and Preibisch**

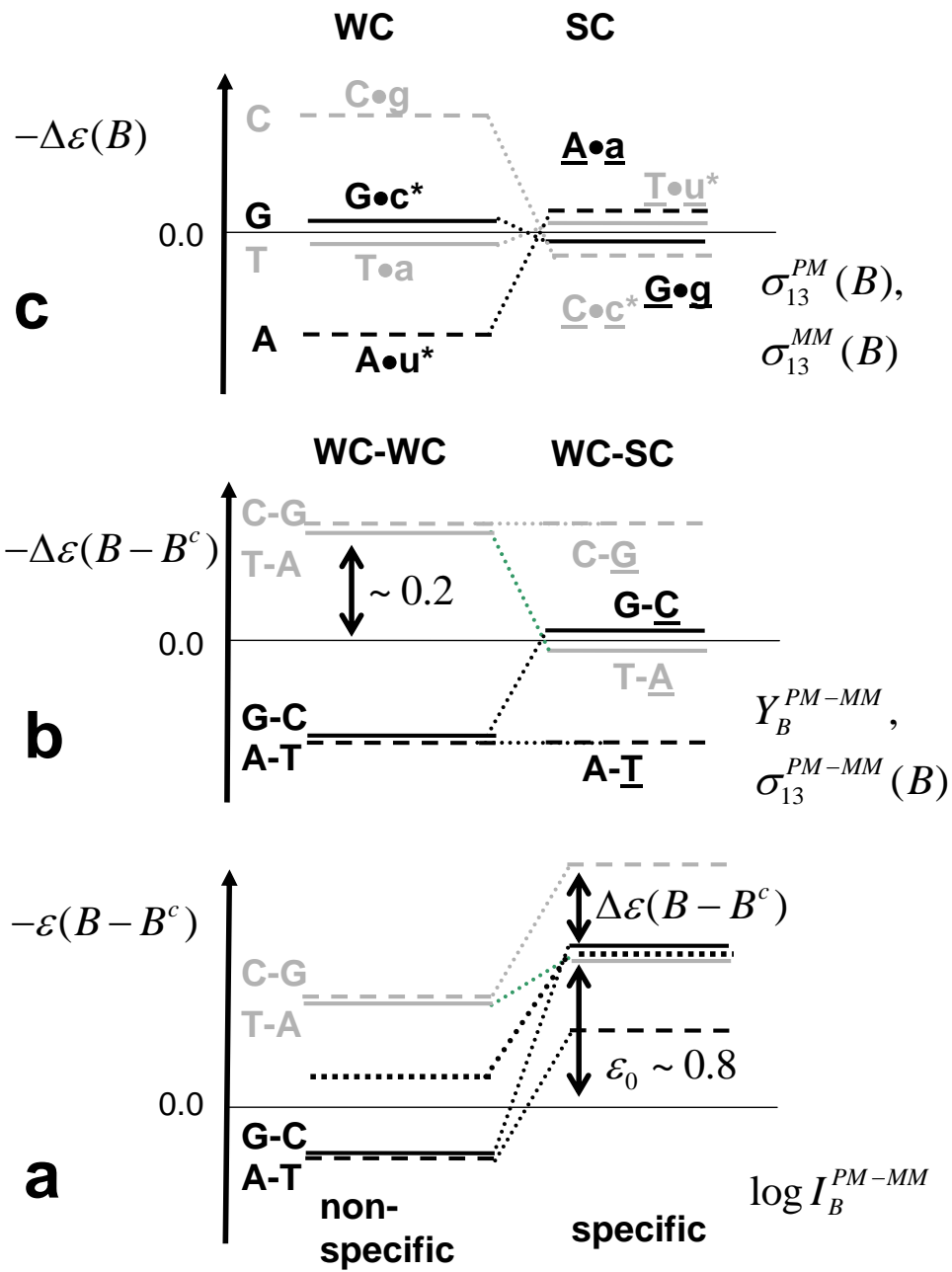

**Figure 11, Binder and Preibisch**

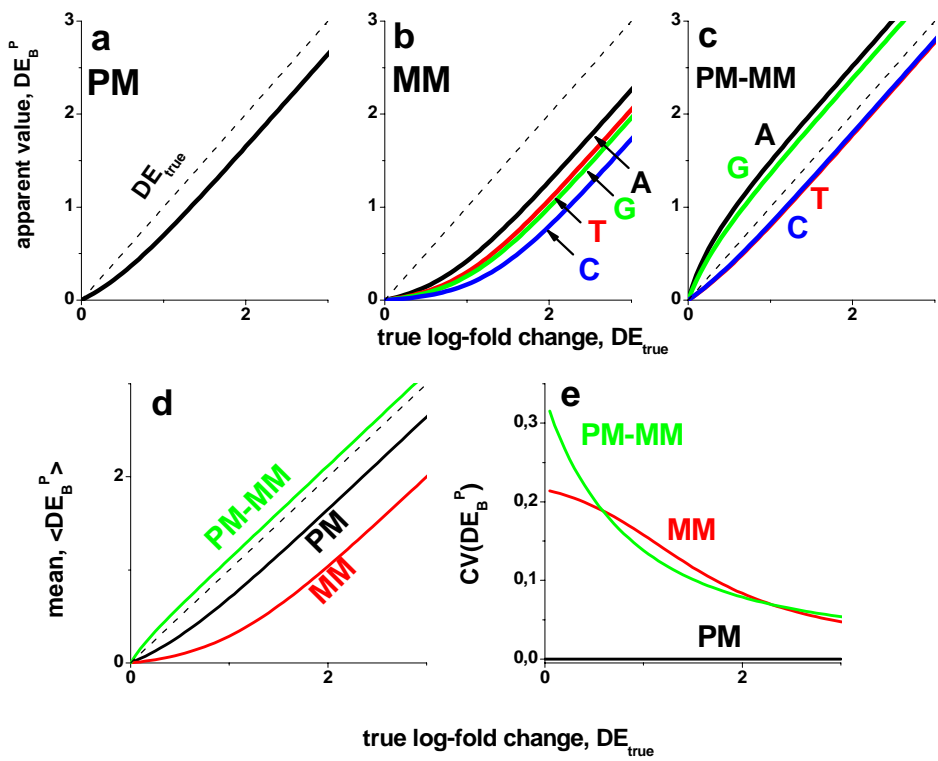

**Figure 12, Binder and Preibisch**